\documentclass[
	a4paper, 
	10pt, 
	unnumberedsections, 
	twoside, 
]{LTJournalArticle}

\runninghead{FREESS: An Educational Superscalar Simulator}

\footertext{\textit{RISC-V Summit Europe, Bologna, 8-12th June 2026}} 

\title{FREESS: A Web-Based Educational Simulator for a RISC-V-Inspired Superscalar Processor with Tomasulo-Style Dynamic Scheduling}

\author{Roberto Giorgi\textsuperscript{1,2} and Miquel Moret\'o Planas\textsuperscript{2}
    \thanks{\tiny Corresponding author: \href{giorgi@unisi.it}{giorgi@unisi.it}.
This project is co-funded by the Ministerio para la Transformación Digital y de la función pública, within the framework of the Plan de Recuperación Transformación y Resiliencia, and by the European Union – NextGenerationEU. The views and opinions expressed are solely those of the author(s) and do not necessarily reflect those of the European Union. Neither the European Union nor the European Commission can be held responsible for them.
    Barcelona Zettascale Laboratory, promoted by the Spanish Ministry for Digital Transformation and the Civil Service, within the framework of the Recovery, Transformation and Resilience Plan - Funded by the European Union - NextGenerationEU
     and via the PNRR M4C2-Inv1.4 Italian Research Center on High-Performance Computing, Big-Data and Quantum Computing, cascade funding project EDGE-ME, MUR-ID: CN0000013.}
}

\date{\footnotesize\textsuperscript{\textbf{1}}Department of Information Engineering and Mathematics, University of Siena, Via Roma 56, 53100 Siena, Italy\\ \textsuperscript{\textbf{2}}Barcelona Supercomputing Center (BSC), Barcelona, Spain}

\renewcommand{\maketitlehookd}{%
	\begin{abstract}
		\noindent
        FREESS (Free Educational Superscalar Simulator) is an open-source teaching environment for instruction-level parallelism in a RISC-V-inspired superscalar processor. It provides a compact, cycle-by-cycle view of register renaming, issue, execution, write-back, commit, and memory ordering in a Tomasulo-style machine. The simulator exposes the register map, free pool, instruction window, reorder buffer, and load/store queues in one textual representation, so the evolution of the hardware state can be followed on screen and reproduced on paper. Runtime parameters such as issue width, queue sizes, and functional-unit latencies can be changed easily, enabling direct comparison among alternative superscalar organizations. The tool has supported Advanced Computer Architecture teaching for about fifteen years and is publicly available on GitHub.
	\end{abstract}
}

\begin{document}

\maketitle 

\section{Educational contribution}

FREESS addresses a recurring teaching difficulty in advanced computer architecture: the complexity of tracking the dynamic scheduling evolution of the hardware structures that implement a superscalar processor. The simulator makes this evolution explicit. Instead of hiding the internal state behind a performance summary, it presents renamed operands, ready-status changes, queue occupancy, structural conflicts, and in-order retirement cycle by cycle. The result is a didactic environment centered on how Tomasulo-style execution really unfolds in hardware \autocite{tomasulo67,Giorgi25-wcae}.

A key contribution is the choice of a compact, textual, and fully synchronized representation of the machine state. This view keeps the causal relation among instructions, resources, and stalls immediately visible. Students can therefore connect dependence chains, buffer pressure, issue bandwidth, and commit constraints without switching among disconnected windows or abstractions. This is especially useful in classroom discussion and in exam-oriented exercises, where the same trace can be mirrored on paper and reasoned about step by step.

\begin{figure}[t]
    \centering
    \includegraphics[width=0.96\linewidth]{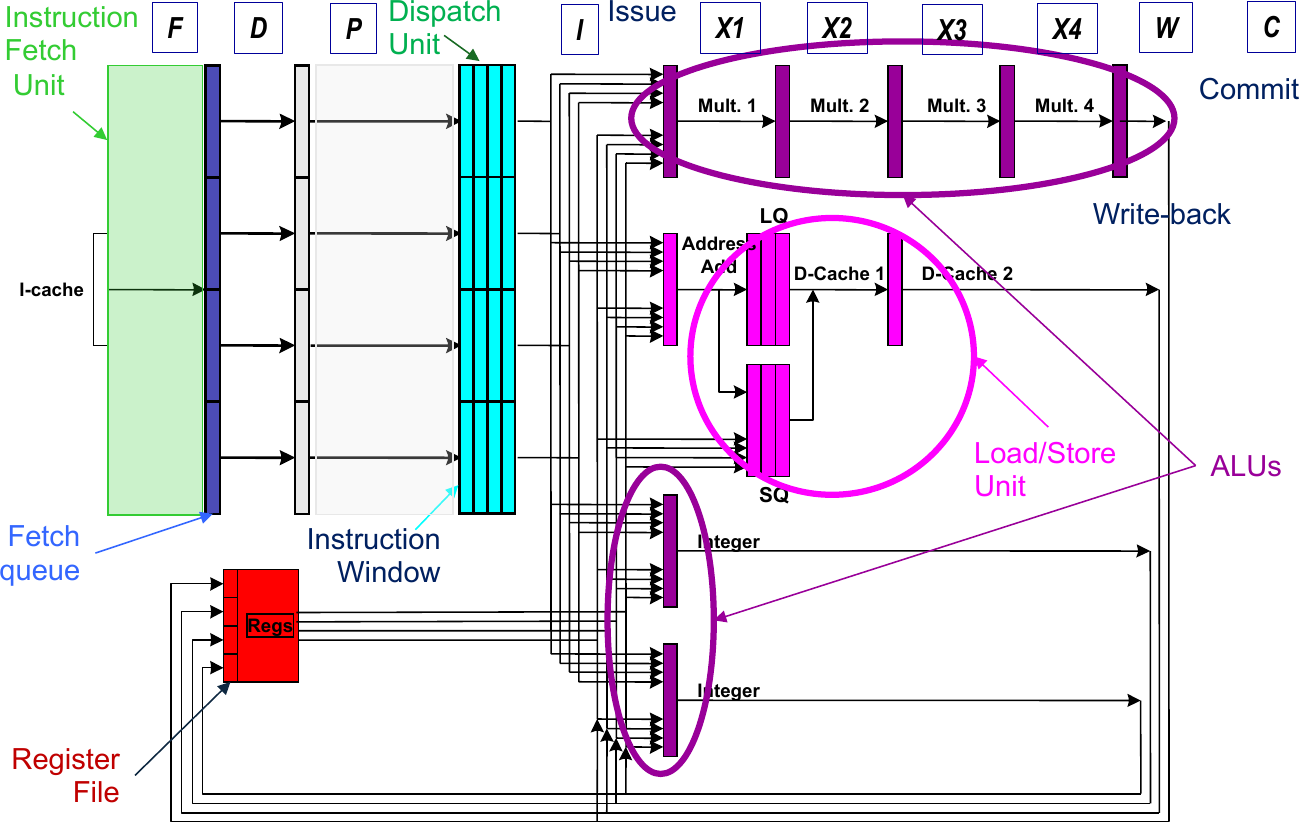}
    \caption{Architectural view used by FREESS to explain the stages and structures of a Tomasulo-based superscalar pipeline.}
    \label{fig:arch}
\end{figure}

\section{Modeled machine and execution environment}

FREESS models a compact superscalar processor around seven instructions: \texttt{ADD}, \texttt{ADDI}, \texttt{BEQ}, \texttt{BNE}, \texttt{LW}, \texttt{MUL}, and \texttt{SW}. The instruction set is intentionally minimal but sufficient to expose the main mechanisms of a superscalar CPU: fetch, renaming, out-of-order dispatch, execution, completion, commit, speculative branching, and memory access. Figure~\ref{fig:arch} summarizes the architectural structures used for this purpose.

The simulator is implemented in C, runs from the command line, and prints a complete cycle-by-cycle trace. The environment is lightweight enough to run easily in a standard teaching setup, yet configurable enough to compare multiple machine organizations. Users can modify superscalar issue width, the number and type of functional units, operation latencies, architectural buffer sizes, queue dimensions, and the number of loop iterations. These controls make it possible to study how a single kernel behaves under different hardware assumptions.

A particularly useful feature is the unified visibility of current state and recent execution history. For each cycle, FREESS shows renamed registers, instruction-window entries, reorder-buffer slots, queue contents, stage occupancy, and the dynamic progress of each instruction. Stall causes are reported explicitly, both in the main trace and in a dedicated log, so structural hazards, unavailable operands, limited issue bandwidth, branch effects, and commit bottlenecks can be discussed directly. Small programs are entered in a simplified machine-code form, which keeps the focus on instruction semantics and microarchitectural behavior rather than on assembler details.

\begin{figure}[b]
    \centering
    \includegraphics[width=0.84\linewidth]{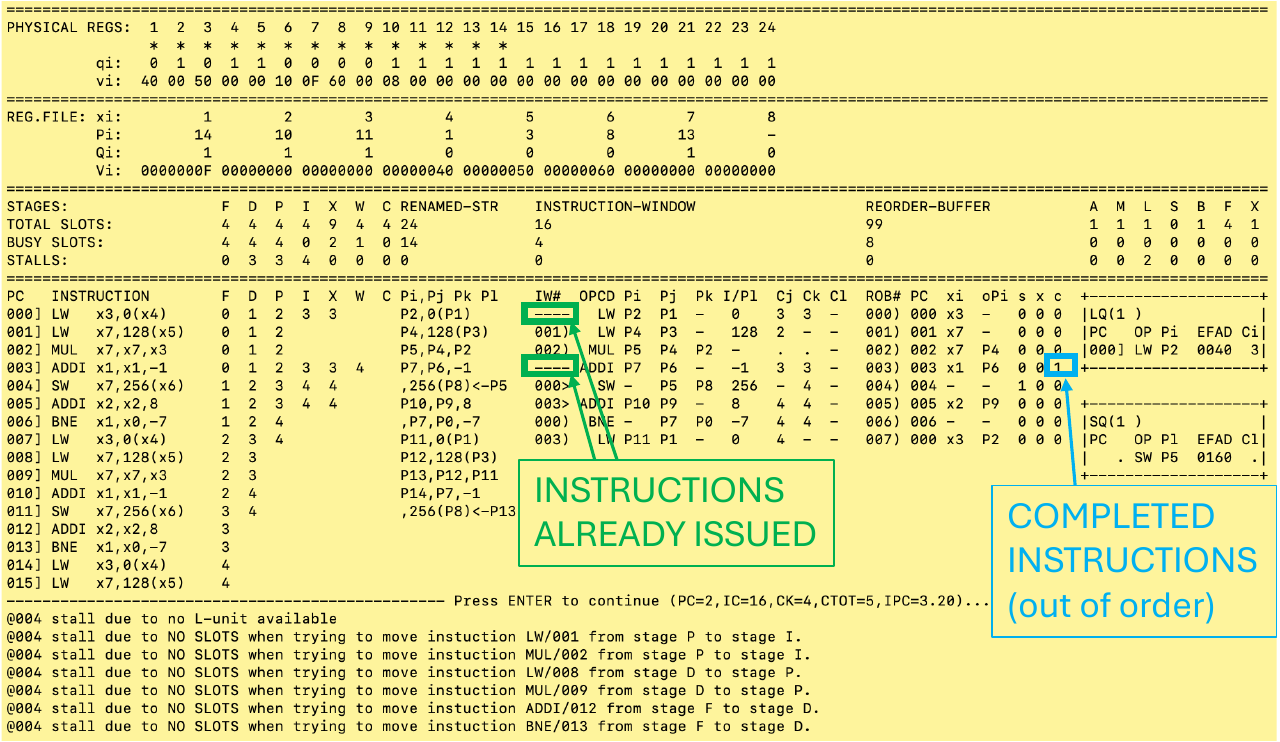}
    \caption{Cycle snapshot showing issued instructions, completed out-of-order operations, instruction-window state, reorder-buffer entries, and stall explanations.}
    \label{fig:snapshot}
\end{figure}

\section{Open source and classroom use}

FREESS is openly released at \url{https://github.com/robgiorgi/freess}. This is more than a distribution detail: the code base is small enough to be inspected and adapted by instructors, while the execution model is simple enough for students to experiment with their own kernels and configurations. The open-source availability therefore extends the tool from a demonstrator to a reusable teaching platform \autocite{Giorgi25-wcae}.

The simulator has been used since 2010, that is, for about fifteen years, in the Advanced Computer Architecture course at the University of Siena. Across many course editions, it has supported lectures on superscalar execution, guided exercises on dynamic scheduling, and discussion of branch behavior. This long deployment is important because it shows that the interface and the modeled machine were refined against real teaching needs.

Compared with educational simulators such as SIMDE and SATSim, and with recent RISC-V-oriented teaching tools, FREESS emphasizes compact visibility of full-machine evolution over richer but more fragmented interaction \autocite{castilla07-simde,Wolff00-satsim,Jaros24-riscv}. That design choice makes it effective for understanding why instructions may complete out of order while still committing in order, and why limits in queues, or functional units translate into observable stalls and throughput changes.

Figure~\ref{fig:snapshot} illustrates this approach. In one screen, students can observe issued instructions, completed operations, queue and buffer state, and the textual explanation of the factors currently limiting progress.

\section{Availability as a Web-Based Tool}

FREESS is provided both as a standalone portable C program and as a web-based
educational tool accessible at \url{https://robgiorgi.github.io/freess/}. This
dual availability improves accessibility in teaching scenarios. The
portable C implementation supports local execution, offline use, and direct
inspection of the source code, whereas the web version enables immediate use
from a browser without installation. Together, these two forms make FREESS both
easy to adopt in classroom demonstrations and flexible enough for hands-on
experimentation by students and instructors.

\section{Conclusions}

FREESS provides a configurable and cycle-accurate educational environment for visualizing Tomasulo-style superscalar execution in a RISC-V-inspired setting. Its main strength is not architectural breadth, but pedagogical clarity: the machine structures that govern dynamic scheduling remain visible, synchronized, and directly traceable throughout execution.

\printbibliography

\end{document}